\documentstyle{article}
\def\Journal#1#2#3#4{{#1} {\bf #2}, #3 (#4)}
\begin{document}
\title{UNIVERSALITY OF EINSTEIN'S GENERAL RELATIVITY}
\author{LESZEK M. SOKO\L{}OWSKI \\ Astronomical Observatory,
Jagellonian University, Orla
171, \\ Krak\'ow 30-244, Poland}
\date{Talk given at the \\14th Conference on General Relativity and
Gravitation\\
Florence (Italy), August 1995}
\maketitle\begin{abstract}
Among relativistic theories of gravitation
the closest ones to general relativity are the scalar-tensor ones
and these with Lagrangians being any function $f(R)$ of the
curvature scalar. A complete chart of relationships between these
theories and general relativity can be delineated. These theories
are mathematically (locally) equivalent to general relativity plus
a minimally coupled self-interacting scalar field. Physically they
describe a massless spin-2 field (graviton) and a spin-0 component of
gravity. It is shown that these theories are either physically
equivalent to general relativity plus the scalar or flat space is
classically unstable (or at least suspected of being  unstable). In
this sense general relativity is universal: it is an isolated point
in the space of gravity theories since small deviations from it either
carry the same physical content as it or give rise to physically
untenable theories.
\end{abstract}

\section{Introduction}
   General relativity is just a point in the "space" of all existing
and conceivable theories of gravitational interactions. Nevertheless
all the theories other than Einstein's one, named "alternative theories
of gravity", have rather bad reputation among most relativists.
General relativity is enough complicated in itself and well confirmed
by all known empirical data so that there is no point in considering
more intricate theories whose empirical basis is, as a rule, either
smaller than that of Einstein's theory or presently non-existing at
all. In fact, the alternative theories of gravity are some
generalizations of general relativity, which invariably serves as a
reference point for constructing them. These modifications go in all
possible directions [1] making their theoretical investigations and
attempts to confront them with experiment so difficult.

   On the other hand in the last years there has been considerable revival
of interest in some alternative theories. If one seeks for a deeper
relationship between gravitational physics and other interactions,
particularly in the realm of elementary particles, then one finds
signals that some modifications of Einstein's theory are inevitable or
at least desirable. In the low-energy field theory limit of superstring
effective action one recovers Einstein-Hilbert Lagrangian plus higher
order corrections in the curvature [2]. Superstring theory gives also
rise to a scalar field, the dilaton, which in this limit is non-minimally
coupled to the string metric [3] and is viewed as a spin-0 partner of
metric gravity. Nonrenormalizability of Einstein's theory and
renormalizability of a quadratic Lagrangian suggests that a classical
limit (to be defined when quantum theory of gravity will become a fact
rather than a fancy) of quantum gravity may be a theory with dynamics
more complicated than that of general relativity. Quadratic and higher
order in Riemann curvature Lagrangians are possible candidates for a
theory avoiding spacetime singularities [4]. Scalar fields are copious
in modern particle theories but in these contexts they do not exhibit
very specific properties needed in the currently popular models of
inflationary
evolution of the early universe; usually one puts an inflaton field in
the theory  just by hand. This is why scalar-tensor theories of gravity,
in which the metric field is supplemented by a scalar having arbitrarily
prescribed features, seem to be very promising, particularly to particle
physicists (and some of them deal with these theories in a rather
careless way) for describing the primordial universe. (Hyper)extended
inflation [5] and kinetic inflation [6] are just few examples. Finally,
and here is a difference to the previous cases, scalar-tensor gravity
theories can be used as test theories probing gravitational physics:
they agree with general relativity in the stationary weak-field limit
(the post-Newtonian approximation) and deviate from it in a strong-field
regime or for radiative fields [7].

   What I have said above does not mean that these (and possibly other)
modifications of general relativity are in a sense superior to it. This
is just an argument for treating them more seriously and a motivation
for investigation of their physical content and relationship to general
relativity. And since GR11 (Stockholm 1986) where a special workshop
session was held on alternative theories of gravitation, a significant
progress has been made and it is possible now to delineate a complete
chart of relationships between scalar-tensor gravity theories, theories
with nonlinear Lagrangians and general relativity.

   Accordingly, I will consider here two modifications of general
relativity: either the Einstein-Hilbert Lagrangian is replaced by an
arbitrary scalar function of the Riemann tensor or there is a spin-0
i.e. scalar-field component of gravitational interaction, which is
non-minimally coupled to the metric.\footnote{These theories are
referred to as {\it metric theories of gravity\/}. Similar techniques
can be applied to purely affine and metric-affine gravity theories,
see a review [8].} A mixture of these two can also be considered and
dealt with in the same way; for simplicity I will omit it here. All
other axioms of general relativity hold for these theories. (By Einstein's
theory I mean general relativity in any dimension $d\geq 4$. I assume
$d=4$ because in Kaluza-Klein theory difficulties arise due to the
existence of multiple ground states corresponding to various topologies
of the extra dimensions while in the case of $d>4$ uncompactified
dimensions the theory is in obvious conflict with experiment. Nonetheless,
formally all the constructions presented below work, with slight
modifications, for any $d\geq 4$.) In this sense the modified theories
form a densely populated neighbourhood of general relativity in the space
of gravity theories. This in turn raises a fundamental question: is
Einstein's theory merely a point of this neighbourhood? As a topology
of the space is undefined, the problem at this level of reasoning has
imprecise, intuitive sense. In other terms: is general relativity
distinguished merely by tradition and computational simplicity or does
it take a preferred position with respect to theories that surround it?
The message of my talk is: {\it these theories are mathematically
equivalent to general relativity and there are convincing arguments that
they are also physically equivalent to it\/}. In a sense these theories
represent Einstein's theory in disguise. General relativity is not
surrounded by theories different from it, its neighbourhood consists of
its own versions in distinct variables. General relativity is an isolated
point in the space of gravity theories. All these notions will be given
a more precise meaning below.

   While investigating these theories it is always assumed that they are
fundamental i.e. independent theories. This means that even if they
arise from other theories (strings, quantum gravity etc.) they are not
subject to rules which are not inherent to them.

   The history of these alternative theories of gravity is long, rich
and begins with the celebrated Weyl's theory in 1918 [9]. I do not
intend to make a survey of this history. Just to give some idea of how
our understanding of relationships of these theories to general relativity
developed in time, I mention here some works. If a paper on the subject
is omitted, it occurs merely due to incompleteness of the list. Special
cases were studied by Higgs 1959 [10], Dicke 1962 [11], Bicknell 1974 [12],
Bekenstein 1974 [13], Stelle 1977--78 [14], Whitt 1984 [15], Barrow and
Cotsakis 1988 [16], Maeda 1989 [17], Schmidt 1990 [18], Cho 1987--93 [19],
Damour, Far\'ese and Nordtvedt 1992--93 [7] and Wands 1993 [20]. General
theory was developed mainly in a series of papers by Magnano, Ferraris
and Francaviglia 1987--90 [21-23], Jakubiec and Kijowski 1988--89 [24]
and Magnano and Soko\l owski 1994 [25]. I will begin with a brief
presentation of what is known in the most general case of nonlinear
Lagrangians.

 \section{Structure of a general metric nonlinear gravity theory}

   A general metric nonlinear gravity (NLG) theory (sometimes named,
rather improperly as will be shown below, "a higher-derivative gravity")
is based on a Lagrangian $L=\sqrt{-g}f(g_{\alpha\beta}, \partial g,
\partial^2g)$ where $f$ is na arbitrary scalar function. Due to
general covariance of the theory, in vacuum the Lagrangian depends only
on the metric and its Riemann tensor, $f=f(g_{\alpha \beta}, R_{\alpha
\beta \mu \nu})$. Except for Einstein-Hilbert Lagrangian and the
Euler-Poincar\'e topological invariant density (Gauss-Bonnet term), the
Lagrangian gives rise to fourth-order field equations.

   It is now both common and useful to refer to the set of dynamical
variables in a gravitational theory as a {\it conformal frame\/} (not to
be confused with the notion of a reference frame); the meaning of
"conformal" will become clear later. In the case of a NLG theory it is
Jordan conformal frame (JCF) and for vacuum theory it consists of the
metric alone, ${\rm JCF}=\lbrace g_{\mu \nu}\rbrace$.

As it is very difficult to study the physical content of a fourth-order
theory one should first lower the order of the field equations. The
best method is to use a canonical Hamiltonian formalism. It should be
stressed that it represents a covariant field-theory version of the
well-known formalism in classical mechanics and it has nothing to do
with the ADM formalism; actually it does not apply to general relativity
at all. In the case of second-order Lagrangians the Hamiltonian
formalism is far from being unique [23]. However in the case of a NLG
theory it is natural to write $L=\sqrt{-g}f(g_{\alpha \beta}, R_{\alpha
\beta}, C_{\alpha \beta \mu \nu})$ and following a purely affine gravity
theory to define a momentum tensor canonically conjugated to the
Christoffel connection as [24]
$$h^{\alpha \beta}\equiv (-g)^{-1/2}\vert {\rm det}({\partial f\over
\partial R_{\mu \nu}})\vert^{-1/2} {\partial f\over \partial R_{\alpha
\beta}}.\eqno(1)$$
(This is in four dimensions, for $d\geq 4$ the exponents are
approprietely altered.) The formalism works if the $10\times 10$ Hessian
$${\rm det}({\partial^2f\over
\partial R_{\alpha \beta} \partial R_{\mu \nu}})\not=0.\eqno(2)$$
Then the definition (1) can be inverted and the Ricci tensor is expressed
in terms of the momenta and all other fields,
$$R_{\alpha \beta}(g)=r_{\alpha \beta}(h^{\mu \nu}, g_{\mu \nu}, C).\eqno
(3)$$
The condition (2) means that the Lagrangian is truly nonlinear in the
curvature. Furthermore, assuming that the momentum is a nonsingular
matrix, i.e. ${\rm det}({\partial f\over \partial R_{\alpha \beta}})
\not=0$, one can view $h_{\alpha \beta}$ as {\it a new spacetime
metric\/}. This corresponds to a mapping from the Lorentzian manifold
$(M, g_{\mu \nu})$ to another one, $(M, h_{\mu \nu})$. The conformal
frame for the gravity theory on $(M, h_{\mu \nu})$, in this case
named Einstein conformal frame (ECF), consists of three independent
fields, ${\rm ECF}=\lbrace h_{\mu \nu}, g_{\mu \nu}, C_{\alpha \beta
\mu \nu}\rbrace$. The tensor fields $g_{\mu \nu}$ and $C_{\alpha \beta
\mu \nu}$ lose their geometrical meaning they had in the initial
spacetime and now can be viewed as a "matter source" for the new
metric; actually they represent additional degrees of freedom for
gravity.

The Legendre transformation (1) is followed in the standard way by
replacing the Lagrangian by a Hamiltonian defined as
$$H(h, g, C)\equiv \sqrt{-h}h^{\mu \nu}r_{\mu \nu}(h, g, C)-
\sqrt{-g}f(g, r, C).\eqno(4)$$
In classical mechanics the canonical Hamilton equations arise as the
variational Euler-Lagrange equations (see section 4) from the
Helmholtz Lagrangian (I use this name following Poincar\'e and
Levi-Civita [26]). In the present case it reads
$$L_H(h, g, C)\equiv \sqrt{-h}h^{\mu \nu}R_{\mu \nu}(g)-H(h, g, C)
\eqno(5)$$
and using an identity relating Ricci tensors for any two distinct
metrics [21], it can be re-expressed, after some manipulations and
discarding a full divergence, in the following form:
$$L_H(h, g, C)=\sqrt{-h}[\overline R(h)+K(h, g)-h^{\mu \nu}r_{\mu
\nu}(h, g, C)]+\sqrt{-g}f(g, r_{\alpha \beta}, C),\eqno(6)$$
where $\overline R(h)$ denotes the curvature scalar for $h_{\mu \nu}$.
This Helmholtz Lagrangian has remarkable properties. Firstly, the free
Lagrangian for the metric is exactly the Einstein-Hilbert one.
Secondly, the kinetic Lagrangian for the "matter" field $g_{\mu \nu}$,
\setcounter{equation}{6}
\begin{eqnarray}
K(h, g) &=& {1\over 2}h^{\mu \nu}g^{\alpha \sigma}g^{\beta \tau}[
\overline\nabla_\alpha g_{\beta \tau} (\overline\nabla_\mu g_{\sigma
\nu}-{1\over 2}\overline\nabla_\sigma g_{\mu \nu})+\nonumber\\& &
+\overline\nabla_\alpha g_{\tau \nu}(\overline\nabla_\sigma g_{\beta
\mu}-\overline\nabla_\beta g_{\sigma \mu})-{1\over 2}\overline\nabla_
\mu g_{\sigma \beta}\overline\nabla_\nu g_{\alpha \tau}]
\end{eqnarray}
is {\it universal\/} i.e. it bears no trace of $f$ [24, 21]
($\overline\nabla
_\mu$ denotes the covariant derivative with respect to $h_{\alpha \beta}
$). Thirdly, the whole information on the original nonlinear Lagrangian
$L$ is encoded in the potential terms depending also on the
$C_{\alpha \beta
\mu \nu}$ field. The variation with respect to $h^{\mu \nu}$ yields
Einstein field equations,
$$\overline G_{\mu \nu}(h)=T_{\mu \nu}(h, g, \overline\nabla g,
\overline\nabla\overline\nabla g, C),\eqno(8)$$
where all the terms in (6) except for $\overline R(h)$ contribute to the
"matter" energy-momentum tensor. It is clear from the form (5) of $L_H$
that these equations are equivalent to the inverse Legendre
transformation, $R_{\mu \nu}(g)=r_{\mu \nu}(h, g, C)$. Independent
variations $\delta g^{\mu \nu}$ and $\delta C_{\alpha \beta \mu \nu}$
provide second-order equations of motion for the fields. Unfortunately,
the equations for $g^{\mu \nu}$ are intractably complex even in the
simplest case of a quadratic $L$ (actually it turns out that the full
complexity of the NLG theory arises already on this level and for
other Lagrangians, polynomial or non-polynomial in the curvature, the
complexity does not substantially increase).
Using the relation $R_{\mu \nu}=
r_{\mu \nu}$ one gets $L_H(h, g, C)=L(g)$, therefore the equations of
motion are equivalent to the fourth-order field equations
${\delta L\over \delta g^{\mu \nu}}=0$ of the original theory (see [24]
for the detailed proof).

Thus there is {\it a dynamical equivalence\/} of any vacuum NLG theory to
general relativity with Lagrangian $L_H(h, g, C)$ for "matter" fields
$g_{\mu \nu}$ and $C_{\alpha \beta \mu \nu}$. This means that although
the action integrals for $L$ and $L_H$ are different in general, their
stationary points (i.e. classical equations of motion) are the same,
what is equivalent to the statement that the spaces of classical
solutions for both theories are isomorphic. In other terms, general
relativity with $L_H$ is {\it a universal Hamiltonian counterpart\/} of
any NLG theory.

As an example let's take the most frequently studied quadratic case
without Weyl tensor, $L=\sqrt{-g}(R+\alpha R^2+\beta R_{\mu \nu}R^
{\mu \nu})$. Here ${\rm JCF}=\lbrace g_{\mu \nu}\rbrace$ while
${\rm ECF}=\lbrace h_{\mu \nu}, \psi_{\mu \nu}, \psi\rbrace$. Using
field propagators in linear approximation [14, 27] one gets a direct
particle interpretation of the fields:
\begin{itemize}
\item $h_{\mu \nu}$ is the massless graviton (spin-2) with 2 degrees of
freedom (d.o.f.),

\item $\psi_{\mu \nu}$ is a massive spin-2 field carrying 5 d.o.f.,

\item $\psi$ is a massive spin-0 particle with 1 d.o.f.
\end{itemize}
The coefficients should satisfy the no-tachyon conditions $m^2_0=
(6\alpha +2\beta)^{-1}>0$ and $m^2_2={-1\over \beta}>0$. The original
metric $g_{\mu \nu}$ is a unifying field carrying together 8 d.o.f. with
different spins.

{}From this example one sees that the mathematically equivalent theories
have different interpretation. While in the NLG theory one views the
massive fields $\psi_{\mu \nu}$ and $\psi$ as finite-range components
of the gravitational interaction, in Einstein's theory one interpretes
them as particular species of elementary particles and describes
gravitation in terms of the metric $h_{\mu \nu}$ alone. Despite the
difference, the particle content of both theories (revealed in ECF) is
the same.

For other Lagrangians, however, hard problems arise. Assume simple purely
quadratic Lagrangian, $L=\sqrt{-g}R_{\mu \nu}R^{\mu \nu}$. Then $h^{\mu
\nu}$ is proportional to $R^{\mu \nu}$ and in the subspace of solutions
for which ${\rm det}(R^{\mu \nu})\not=0$ the tensor $h^{\mu \nu}$ in
general does not have Lorentz signature and thus cannot describe the
physical spacetime metric. Furthermore, the non-geometric components
of gravity, $g_{\mu \nu}$ and $C_{\alpha \beta \mu \nu}
$, are formally
interpreted in ECF as sources of metric gravity, i.e. as matter fields.
What can then be said about their energy? In the light of the well-known
inconsistency of minimal coupling of the linear spin-2 field to geometry
[28] one expects that a physical theory of these fields, although free
of inconsistencies (since they are absent in JCF they cannot arise in
any frame dynamically equivalent to it), will not be easy to formulate
(e.g. it is known that the tensor field is a ghost). These are open
questions and at the present level of art it is prudent to say that for
a generic NLG theory its mathematical equivalence to general relativity
(including the "matter" fields) needs not to imply their physical
equivalence.

In what follows I will confine myself to the restricted NLG theories,
where these problems do not arise. It is convenient to consider first the
scalar-tensor theories of gravity.

\section{ Scalar-tensor gravity}

These are theories in which gravitational interactions are described by
a doublet consisting of a spacetime metric and a scalar field. It is here
that the notion of Jordan conformal frame was introduced [29]. Thus
${\rm JCF}=\lbrace g_{\mu \nu}, \varphi\rbrace$, we consider first a
vacuum theory. It was first recognized by Pauli in early fifties (quoted
in Sect. 28 of ref. [30]) that in such a system one can always make a
field redefinition (a change of variables) to another conformal frame via
a conformal mapping $g_{\mu \nu}\rightarrow \tilde g_{\mu \nu}=\Omega^2
(\varphi)g_{\mu \nu}$ and $\varphi\rightarrow \tilde \varphi=
\tilde \varphi(\varphi)$ with arbitrary $\Omega$ and $\tilde \varphi$.
Thus the theory can be expressed in terms of infinite number of conformally
related frames. A general Lagrangian for a scalar-tensor gravity (STG)
theory,
$$L=\sqrt{-g}[\varphi R-{\omega(\varphi)\over \varphi}g^{\mu \nu}
\varphi_{,\mu}\varphi_{,\nu}+2\varphi V(\varphi)],$$
can then take, among others, the equivalent forms
$$L=\sqrt{-\tilde g}[f(\tilde \varphi)\tilde R+\tilde V(\tilde\varphi)]=
\sqrt{-\bar g}[\bar R(\bar g)-\bar g^{\mu \nu}\phi_{,\mu}\phi_{,\nu}-
\bar V(\phi)].$$
The first of these contains {\it no\/} kinetic term for $\tilde\varphi$ and
a propagation equation for the scalar arises due to the nonminimal
coupling to the curvature. The other form represents just a
self-interacting scalar $\phi$ minimally coupled to gravity and is
designated as Einstein conformal frame, ${\rm ECF}=\lbrace \bar g_{\mu
\nu}, \phi\rbrace$, with $\bar g_{\mu \nu}=\varphi g_{\mu \nu}$ and
$d\phi\equiv [\omega(\varphi)+{3\over 2}]^{1/2}{d\varphi \over
\varphi}$ for $\omega >-{3\over 2}$. It should be stressed that these two
forms are {\it not\/} special cases of the general Lagrangian, but are
equivalent to it for any $\omega(\varphi)$ [25, 27].

Thus apparently different STG theories can be mapped onto each other
by conformal mappings. Consider for simplicity the case without
self-interaction, $V(\varphi)=0$. Then all the theories are divided
in two classes: those with $\omega=0$ (this class actually contains
only one member) and with $\omega\not=0$. The conformal transformations
map theories within the classes and from one class to the other, the
potential is always zero. Each STG theory can be transformed into
general relativity plus conformally invariant scalar field $\chi$ [25].
The latter is usually interpreted as different from a STG theory: the
conformally invariant scalar is commonly viewed as a special kind of
matter rather than being a spin-0 component of gravity; in early seventies
it was believed that the field would exhibit more interesting features
in quantum theory than the ordinary scalar. Despite the traditional
interpretation, the scalar $\chi$ fits the general framework of STG
theories. It had been discovered by Bekenstein [13] and remained
unnoticed for many years that the conformally invariant field is
equivalent under a conformal map to the massless linear scalar minimally
coupled to Einstein gravity. Now it is known that this is merely a
special case of a generic feature: each STG theory is equivalent to
general relativity plus the massless linear scalar field [25]. In other
terms this ordinary scalar may be represented in disguise in infinite
number of ways as the spin-0 gravity component in any STG theory or as the
conformally invariant field.

What about interactions with matter? Among all conformally related frames
one distinguishes two frames: JCF and ECF. As scalar fields have not yet
been observed in nature, one can in principle assume any form of their
interaction with ordinary matter (here collectively denoted by $\psi$).
Actually only two possibilities seem to be physically interesting and
reasonable.

1. Matter minimally couples to the metric in JCF,
$$L(g, \varphi, \psi)={1\over 16 \pi}(\varphi R-{\omega\over \varphi}
g^{\mu \nu}\varphi_{,\mu}\varphi_{,\nu})+L_{\rm mat}(g, \psi),$$
"to the only physically meaningful frame whenever one has to deal with
a Jordan-Brans-Dicke-type theory" (S. Matarrese).

2. Minimal coupling in ECF,
$$L={1\over 16\pi}[\bar R(\bar g)-\bar g^{\mu \nu}\phi_{,\mu}\phi_{,\nu}]
+L_{\rm mat}(\bar g, \psi).$$
One {\it first\/} makes the conformal transformation to ECF and {\it
then\/} couples matter to metric gravity in it. Clearly no trace of the
original STG theory survives in this frame and most advocates of these
theories reject this form of coupling.

Once matter has been coupled to gravity in a frame one has freedom to make
conformal transformations to any other frame. E.g., if matter is
minimally coupled in JCF (version 1) then the theory is described in ECF
by
$$L'=\sqrt{-\bar g}\left[\bar R(\bar g)-\bar g^{\mu \nu}\phi_{,\mu}
\phi_{,\nu}+
\left({16\pi\over \varphi(\phi)}\right)^2L_{\rm mat}\left({16\pi\over
\varphi(\phi)}
\bar g, \psi\right)\right];$$
in this frame the scalar directly interacts with matter. Similar
interactions arise when one transforms the theory with matter minimally
coupled in ECF (version 2) back to JCF. The two versions of coupling
matter to gravity generate two {\it physically different\/} gravity
theories. Thei difference lies in distinct effects they predict for
matter while it should be emphasized that each theory is internally
consistent in any frame; e.g. conservation laws hold for them in all
frames [25].

\section{Restricted nonlinear gravity theories}

These are metric gravity theories in which the Lagrangian depends on the
Riemann tensor solely via the curvature scalar. In vacuum $L=\sqrt{-g}
f(R)$ and ${\rm JCF}=\lbrace g_{\mu \nu}\rbrace$, where $f$ is any smooth
function (later I shall assume that $f$ is analytic around $R=0$).
Except for $f=R$ the field equations are of fourth order. To lower their
order one cannot use exactly the same canonical formalism as in the
general case since the Legendre map is degenerate,
$${\rm det}({\partial^2f\over \partial R_{\alpha \beta}\partial R_{\mu
\nu}})=0.$$
One introduces instead a scalar momentum canonically conjugated to a
linear combination of connection,
$$p\equiv {1\over \sqrt{-g}}{\partial L\over \partial R}=f'(R).\eqno(9)$$
The regularity conditions are then $f'(R)>0$ and $f''(R)\not=0$. Then
the definition (9) can be inverted to yield $R(g)=r(p)$, e.g. for
$f=R+aR^2$ there is $r(p)={1\over 2a}(p-1)$. The field Hamiltonian is then
$$H(p, g)\equiv \sqrt{-g}[pr(p)-f(r(p))].\eqno(10)$$
The Hamilton equations for a mechanical system arise as stationary points
of the action for the Helmholtz Lagrangian defined as a function on the
tangent bundle to the phase space:
$$L_H(q, p, \dot q, \dot p)\equiv p\dot q-H(q, p).\eqno(11)$$
In fact, the independent variations $\delta q$ and $\delta p$ of the
action yield correspondingly
$$\dot p=-{\partial H\over \partial q}\quad {\rm and}\quad \dot q=
{\partial H\over \partial p}.$$
The Helmholtz Lagrangian for a NLG theory reads
$$L_H(p, g)=\equiv \sqrt{-g}\ pR(g)-H(p, g).\eqno(12)$$
Since p is an independent degree of freedom we are now working in
Helmholtz-Jordan conformal frame (HJCF) consisting of $g_{\mu \nu}$ and
$p$. The Hamilton equations following from $L_H$ are of second order,
\setcounter{equation}{12}
\begin{eqnarray}
{\delta L_H\over \delta p}=0 &\Rightarrow& R(g)=r(p),\\
{\delta L_H\over \delta g^{\alpha \beta}}=0 &\Rightarrow&
G_{\alpha \beta} = \begin{array}[t]{l}
{1\over p}(\nabla_\alpha\nabla_\beta p-g_{\alpha \beta}
\nabla_\mu\nabla^\mu p)\\
\\-{1\over 2}[r(p)-{1\over p}f(r(p))]g_{\alpha \beta}
\equiv \theta_{\alpha \beta}(g, p).\end{array}
\end{eqnarray}
By introducing the scalar momentum one not only reduces the fourth-order
field equations to second-order ones but moreover these are precisely
Einstein equations with the momentum generating a "matter source". I stress
that this has been achieved without altering the spacetime metric [23, 25]
and the theory is not inherently higher-derivative one.

Investigation of the Cauchy problem for $L=\sqrt{-g}f(R)$ shows that the
scalar $p$ represents an independent dynamical degree of freedom [31].

The NLG theory with the Lagrangian $L$ in JCF is dynamically equivalent to
the theory with Helmholtz Lagrangian $L_H$ expressed in HJCF. The latter
describes 3 d.o.f., hence the metric $g_{\mu \nu}$ in JCF unifies spin 2
and spin 0. This is a special case of the general rule: the number of spin
d.o.f. carried by a field depends on its tensorial character and the
equations of motion it satisfies. If the metric Lagrangian does not contain
explicitly the Ricci tensor, the number of d.o.f. decreases from 8 to 3.

The Helmholtz Lagrangian (12) describes a STG theory in a frame where the
kinetic part of the Lagrangian for the scalar $p$ is absent, $\omega=0$.
Thus the restricted NLG theory with $L=\sqrt{-g}f(R)$ is equivalent to an
STG theory with a self-interacting scalar field. The latter theory can, as
we have seen, be transformed into general relativity with a self-interacting
minimally coupled scalar field $\phi$. The conclusion is that any
restricted vacuum NLG theory is dynamically equivalent to the nonlinear
scalar $\phi$ in Einstein's gravity theory. The equivalence is attained in
two steps. First one maps a given NLG theory via the Legendre map into a
STG theory and then the latter is transformed with the aid of a conformal
change of variables, $\bar g_{\alpha \beta}\equiv pg_{\alpha \beta}$,
$\phi \equiv \sqrt{3\over 2}\ln p$, into general relativity plus $\phi$
with
$$L_H(\bar g, \phi)=\sqrt{-\bar g}[\bar R(\bar g)-\bar g^{\alpha \beta}
\phi_{,\alpha}\phi_{,\beta}-2\bar V(\phi)],\eqno(15)$$
where the potential is
$$\bar V(\phi)={1\over 2}({r(p)\over p}-{f(r)\over p^2})\eqno(16)$$
and ${\rm ECF}=\lbrace \bar g_{\mu \nu}, \phi \rbrace$.

The equivalence means that the conformal transformation and Legendre map
can (at least in principle) be inverted. In fact, for any
self-interaction potential $\bar V(\phi)\not=0$ in ECF the {\it inverse
problem of nonlinear gravity\/} has a solution [25]: there exists a
vacuum ${\rm JCF}=\lbrace g_{\mu \nu}\rbrace$ and a Lagrangian $L=
\sqrt{-g}f(R)$ in it such that $L$ is equivalent to $L_H$ given in (15).
In most cases, however, $L$ cannot be expressed in terms of elementary
functions. One of few exceptions is provided by the Liouville field theory
with $\bar V(\phi)=A \exp (\sqrt{2\over 3}\phi)$, then $f(R)=
4A(6AR)^{3/2}$. The linear massless ($\bar V=0$) field $\phi$ is not
equivalent to a restricted NLG theory but to a STG theory. It is worth
to emphasize the difference between the two steps. STG theories are
equivalent to Einstein's theory plus the scalar in the sense of equality
of the action integrals for the theories, thus the equivalence holds not
only for the solutions of the classical field equations. Yet for NLG
theories on one hand and for STG theories and general relativity on the
other, the action integrals are different in general and these are the
spaces of classical solutions that are isomorphic. (Actually the isomorphism
is usually local since the Legendre map is only locally invertible, I will
not discuss here this difficult problem.)

When ordinary matter is taken into account the same problem appears as in
the case of STG theories: to which metric should it be minimally coupled?
If it is minimally coupled in JCF,
$$L=\sqrt{-g}[f(R)+2L_{\rm mat}(g, \psi)],\eqno(17)$$
then after making the Legendre map and conformally transforming it takes
on the form in ECF:
$$L_H(\bar g, \phi, \psi)=\sqrt{-\bar g}[\bar R(\bar g)-\bar g^{\mu \nu}
\phi_{,\mu}\phi_{,\nu}-2\bar V(\phi)+2e^{-2\sqrt{2/3}\phi}L_{\rm mat}
(e^{-\sqrt{2/3}\phi}\bar g, \psi)].\eqno(18)$$
On the other hand, if matter is minimally coupled in ECF,
$$L_H=\sqrt{-\bar g}[\bar R(\bar g)-\bar g^{\mu \nu}\phi_{,\mu}\phi_{,\nu}+
2L_{\rm mat}(\bar g, \psi)],\eqno(19)$$
one can make a Legendre transformation to absorb the scalar into the
metric field and obtain again a NLG theory, this time in the presence of
matter. Contrary to a naive view the Legendre map needed to this aim is
not the inverse of the map from the original JCF to ECF. In other words
one has an open chain of frame mappings: $\lbrace g_{\mu \nu}\rbrace
\;\rightarrow\;\lbrace \bar g_{\mu \nu}, \phi\rbrace\;\rightarrow\;
\lbrace \bar g_{\mu \nu}, \phi, \psi\rbrace\;\rightarrow\;\lbrace
\hat g_{\mu \nu}, \psi\rbrace$ with $g_{\mu \nu}\not=\hat g_{\mu \nu}$;
interaction with matter results in an appropriate change of the Jordan
frame metric [25]. In matter Jordan conformal frame $\lbrace \hat g_
{\mu \nu}, \psi\rbrace$ the metric and matter variables are inextricably
intertwined in the resulting nonlinear Lagrangian.

The two ways of coupling with matter give rise to two physically distinct
gravity theories. Contrary to some claims (there was a debate in Phys.
Rev. D, 1995) {\it both theories are consistent in any frame\/} [25]. The
fundamental problem is then: which frame (if any) contains the spacetime
metric of the physical world? Is the problem meaningful at all?

\section{Energy and the choice of a physical frame}

Physical laws are not conformally invariant and properties of elementary
particles are altered under a conformal map. The most general argument
regarding particle masses is provided by quantum mechanics. Under a
conformal rescaling of the spacetime metric, $g_{\mu \nu}\mapsto
\Omega^2g_{\mu \nu}$, the particle's wave function transforms as
$\psi\mapsto\Omega^{-3/2}\psi$. The covariant momentum operator
$p_\mu=-i{\partial \over \partial x^\mu}$ is independent of the metric.
The momentum eigenfunctions of a free particle should satisfy the
equation $p^\mu \psi p_\mu \psi=m^2\psi^2$ both in the original metric
and in the rescaled one. Assuming that the conformal factor $\Omega$
varies slowly on the distances of the order of the particle's Compton
wavelength, one finds that the particle masses scale uniformly under the
conformal mapping: $m\mapsto\Omega^{-1}m$. (It is amusing to
notice that some authors prove this relation in the flat cosmological
Friedmann model and derive from it far-reaching conclusions.)

Some advocates of STG and restricted NLG theories claim that Jordan frame
is physical since the physical frame is the one in which "atomic masses
are constant". In fact, if matter is minimally coupled in JCF as is done
in most papers on the subject, then in ECF particle masses are
spacetime-dependent. It is clear however from what I have said above that
this argument is based on the {\it petitio principii\/} error. Assuming
that in JCF the Lagrangian has the form (17) amounts to assuming that this
frame is physical in the common sense of the term and the argument is a
trivial check of internal consistency of the assumption. One can equally
well assume that in ECF the correct form of the Lagrangian is given by
(19) and then the particle masses are physical constants in this frame
while in matter Jordan conf. frame the highly nonminimal coupling makes
the masses variable. The problem of which frame is physical cannot be
solved in this way.

The problem should be somehow addressed by anyone who deals with STG and
NLG theories. There are posssible four answers to the problem and all are
found in the current literature. The first is that JCF is physical
while ECF is merely a useful computational tool; this
view is shared by most authors applying these theories to various problems
in cosmology and high energy physics. The second is that ECF is physical,
the motivation is less obvious and various authors in this group use
different arguments to support the view: (i) ADM Hamiltonian formalism
works in this frame and ADM and Bondi-Sachs masses for an isolated
system are well defined; (ii) quantization of scalar field fluctuations
should be done in ECF while in JCF the procedure is at least suspect
("it appears as if the quantization and a conformal transformation are
two mutually noncommutable procedures"); (iii) the massless spin-2
graviton is described by the difference $\bar g_{\mu \nu}-\eta_{\mu \nu}$
while in JCF the fluctuations of $g_{\mu \nu}$ about the flat space
represent a mixture of the tensor and scalar fields; (iiii) ECF is
singled out by dimensional reduction of higher-dimensional Einstein's
gravity. Authors in the third group claim that classically JCF and ECF
are physically equivalent: since mass ratios of elementary particles
are unchanged by conformal transformations then "physics cannot
distinguish between conformal frames". They seem to overlook that other
physical quantities, notably energy, are sensitive to these
transformations. The fourth answer, implicitly contained in some works
is just to avoid addressing the problem (see [25] for references). All
authors admit that ECF is {\it always\/} (except for very special
cases) computationally advantageous.

The problem should be ultimately solved by experiment but it is clear
that we are very far from it. On purely theoretical grounds one can
provide convincing arguments in favour of Einstein frame.

Consider a nonrelativistic charged particle in an external electromagnetic
field and perform in its phase space the canonical transformation
renaming the positions and momenta: $P_i=-x_i$ and $Q_i=p_i$ ($x_i$
being Cartesian coordinates). Then its Hamiltonian reads
$$H={1\over 2m}\sum_i [Q_i-{e\over c}A_i(-P_k, t)]^2+eA_0(-P_k,
t)$$
and the rule of minimal coupling to electromagnetic field, symbolically
expressed as ${\bf p\rightarrow p}-{e\over c}{\bf A}$, clearly does
{\it not\/} hold. The rule applies to the momenta canonically conjugated
to the variables on which the field depends. These variables---the
physical coordinates---can be determined in the absence of the field by
requiring that the Hamiltonian of the free particle be independent of
them. Due to homogeneity of the space and time the Hamiltonian should
depend only on the particle's momentum.

In classical general relativity all observable quantities can be
determined by taking into account the physical nature and gravitational
interactions of material bodies forming the reference frame (see e.g.
[32]). This seems to undermine our conjecture that the correct physical
metric can be determined for a vacuum theory. However this is not so since
the role of a material reference frame is here played by the scalar field
which inevitably arises in the gravity theories we consider. In a sense
the scalar field is analogous to the electromagnetic field in the
example and the spacetime homogeneity is replaced by total energy (the
scalar cannot be, however, "switched off" ). In fact, energy plays a
distinguished role in gravitational physics being effectively a conserved
(nonnegative) charge. In this respect the scalar should not
substantially differ from known (classical) matter.

Accordingly, the crucial conjecture is that the physical metric should
be singled out for {\it vacuum\/} theory and this should be done by
considering the interaction of a given metric with the scalar field.
Matter is minimally coupled to the physical metric. Then the form of
the coupling in any conformal frame determined by the relation of the
metric in this frame to the physical metric. The physical metric need
not exist for any gravity theory; actually it exists only for physically
viable theories. A theory of gravity is {\it physical\/} if there exists
a classically stable maximally symmetric ground state solution for it.
(A theory may have several ground states and some of them can be
semiclassically unstable.) A physical theory can be expressed in terms of
various conformal frames of which only few are physical. A conformal
frame is physical if the dynamical variables constituting it are (at
least in principle) measurable and their fluctuations around the ground
state solution have {\it positive\/} energy. In other terms the ground
state solution represents the minimum of energy expressed in physical
variables. We shall employ the well known relation between stability
and positivity of energy. (In practice one does not compute the total
energy of the system; a given solution is stable if any perturbations
have positive energy density. In a theory of gravity expressed in an
unphysical frame the energy density for a fluctuation is indefinite in
general and the relationship to stability is broken.)

Let $$L=\sqrt{-g}f(R)=\sqrt{-g}(R+aR^2+bR^3+\cdots),\quad a\not=0
\eqno(20)$$
and $\Lambda=0$. Then $g_{\mu \nu}=\eta_{\mu \nu}=\bar g_{\mu \nu}$ is
a candidate ground state solution. Near Minkowski space i.e. for
$R\approx 0$ one finds $p=f'(R)=1+2aR+\cdots >0$ and $f''=2a+\cdots
\not=0$, thus the Legendre map is invertible and the metric $\bar g_{\mu
\nu}$ has the correct signature. This shows the local (in the vicinity
of flat space) equivalence of Jordan and Einstein frames. Furthermore
let the spacetime $(M,g_{\mu \nu}$ be asymptotically flat,
$$g_{\mu \nu}=\eta_{\mu \nu}+O(r^{-1/2-\epsilon}),\quad g_{\mu \nu,\alpha}
=O(r^{-3/2-\epsilon}), \quad \epsilon>0.\eqno(21)$$
Then {\it the positive energy theorem for NLG theories\/} holds [25].
Let $\Sigma$ be an asymptotically flat, nonsingular spacelike
hypersurface in a spacetime $(M, g_{\mu \nu})$ topologically equivalent
to $I\!\! R^4$. If (i) the Lagrangian is given by (20), (ii) $p>0$ and
$f''(R)\not=0$ everywhere on $\Sigma$, (iii) a solution $(g_{\mu \nu},
p)$ in HJCF to the field equations (13)--(14) satisfies the condition
(21) and (iv) the coefficient $a>0$, then
a) the potential $\bar V(\phi)$ given by (16) is non-negative on $\Sigma$
and

b) the ADM energy formally defined in JCF in the same way as in general
relativity, is non-negative,
$$E_{ADM}[g]={1\over 2}\int_{S^2}dS_i(g_{ij,j}-g_{jj,i})\geq 0\eqno(22)$$
and vanishes only in flat space.

One sees that under the assumptions of the theorem Minkowski space is a
classically stable ground state solution in both Jordan and Einstein
frames. The proof is an extension and modification of that given by
Strominger [33] for $f=R+a^2R^2$.
As in the case of the classical Positive Energy Theorem in general
relativity , the proof is based on the dominant energy condition for the
source, in this case for the scalar. Therefore the proof goes only in ECF
where $\phi$ is minimally coupled to $\bar g_{\mu \nu}$ and its potential
$\bar V(\phi)\geq 0$. It should be stressed that although the positive
energy theorem does hold in JCF (more precisely, in HJCF), it cannot be
proven in this frame, the existence of ECF is essential. Energy is well
defined for systems described by second-order equations of motion and
these are achieved in HJCF. In this frame $E_{ADM}$ is equal to energy in
ECF. Equality of the total energy in these two frames reflects the fact
that (being a charge) it is evaluated at spatial infinity (where $p
\rightarrow 1$) and is rather loosely related to the interior of the
system. The only detailed information about the interior that is needed is
whether all local energy flows are timelike or null (dominant energy
condition). This connection is lost in JCF. The energy-momentum tensor
$\theta_{\alpha \beta}(g, p)$ for $p$, defined by (14), is indefinite and
{\it a priori\/} negative energy density and superluminal energy flows
may occur. These flaws do not reflect the genuine properties of spin-0
gravity and do not imply that the scalar particles are tachyons; these
are merely due to an improper choice of field variables. $\theta_{\alpha
\beta}$ is {it not\/} the physical energy and momentum density for $p$.
Any field redefinition of the scalar will not help and only with the aid
of the conformal rescaling of the metric one finds the correct expression
for these quantities.

Our conclusions regarding the features of a physical restricted NLG theory
are following.

1. Its Lagrangian must contain the linear term $R$. It ensures that the
Legendre map to ECF exists near Minkowski space (i.e. for $R\approx 0$)
which is supposed to be a ground state solution.

2. $L$ should contain the quadratic term $aR^2$. It ensures regularity
(invertibility) of the Legendre map at flat space.

3. The coefficient $a$ determines stability of flat space. For $a>0$
Minkowski space is a stable ground state solution. For $a<0$ the potential
$\bar V(\phi)$ attains maximum for flat space and renders it classically
unstable. Existence of another solution with minimum of energy is unclear.

4. If a theory is physical then Einstein frame meets all general
requirements of relativistic field theories and is regarded as {\it
physical\/}. The metric $\bar g_{\mu \nu}$ of this frame determines the
spacetime intervals in the physical world. It is related to the original
metric variable of JCF by $\bar g_{\mu \nu}=f'(R)g_{\mu \nu}$. On the
contrary, Jordan frame is never physical: its variables do not provide
a meaningful and tractable relationship to the total energy of the
system.

5. The transformation from JCF to ECF is physically interpreted as a
transition to dynamical variables describing fields with definite spins
and for which the local energy flows are causal implying the positive
energy theorem. There is a geometrical analogy: ${\rm ECF}=\lbrace
\bar g_{\mu \nu}, \phi\rbrace$ is like Cartesian coordinates for
dynamical variables while JCF and other frames are a kind of
"curvilinear coordinates" which can generate fictitious "coordinate
singularities".

Whenever Einstein frame does not exist in the vicinity of flat space,
the theory is either unphysical, e.g. for $f=R^2$ [12], or is suspected
of being such and showing that it is a viable one is difficult.

Final conclusion is that a physical restricted NLG theory is nothing but
general relativity plus the scalar field in a disguise. Such a theory
cannot provide new physical effects different from those existing in
Einstein's theory. If there is a deeper motivation for considering
NLG theories (e.g. as being a field-theory limit of string theory) then
the function $f(R)$ generates the potential $\bar V(\phi)$ for the
scalar field.
\section*{Acknowledgments}
I am deeply grateful to Marco Ferraris, Mauro Francaviglia and my
coauthor, Guido Magnano, for helpful comments and discussions. I
acknowledge the hospitality of "J.-L. Lagrange" Institute of
Mathematical Physics at Torino where a part of this work was done.
This work was partially supported by a grant of Polish Committee for
Scientific Research.

\end{document}